\documentclass[twocolumn]
{openjournal}

\usepackage{upgreek}

\begin{document}

\title{{The orbital parameters
of the binary BLAP HD\,133729\,:\\
advantages of the frequency modulation method}}

\author{Hiromoto Shibahashi}
\affiliation{Department of Astronomy, University of Tokyo\\
Bunkyo-ku, Tokyo 113-0014, Japan}

\author{Donald W. Kurtz}
\affiliation{Centre for Space Research, North-West University\\
Dr Albert Luthuli Drive, Mahikeng 2735, South Africa}
\affiliation{Jeremiah Horrocks Institute, University of Central Lancashire,\\
Preston PR1 2HE, UK}

\author{Simon J. Murphy}
\affiliation{Centre for Astrophysics, University of Southern Queensland\\
Toowoomba, QLD 4350, Australia}

\begin{abstract}
We derive all the orbital parameters of the blue large-amplitude pulsator (BLAP) in the binary system HD\,133729 by exploiting the frequency modulation (FM) method, which is based on the analytical relations between the orbital parameters and a multiplet separated by the orbital frequency in the frequency spectrum of the light curve. Because the FM method uses the entire data through the Fourier transform, it is the most effective use of high-precision photometry data, taken over a long timespan by the TESS space mission, for determining orbital parameters.    
\end{abstract}


\section{Introduction}
Blue large-amplitude pulsators (BLAPs) are a recently discovered class of pulsating variables with effective temperatures of $\sim$$30\,000\,{\rm K}$ \citep{Pietrukowicz2017}. 
The total number of identified BLAPs already exceeds 80 \citep{Pietrukowicz2024}. 
Their typical pulsation periods are of the order of a few tens of minutes, and with relatively large amplitudes of $\sim$$0.3\,{\rm mag}$ in optical bands. They lie in a region of the HR diagram between the upper main sequence and subdwarf B stars. However, the evolutionary origin and current status are yet unclear, mainly due to a lack of reliable mass constraints.

Among the  identified BLAPs, HD\,133729 is the first that has been found to be in a binary \citep{Pigulski2022}. 
This system consists of a BLAP and a B main-sequence star \citep{Pigulski2022}. 
As a pulsating star moves in its binary orbit, the path length of the light between us and the star varies, leading to 
a periodic variation in the arrival time of the signal at the solar system barycenter and, as a consequence, 
pulsations of a star in a binary system show phase modulation with the orbital frequency due to the light travel time effect 
\citep{PASP335,Murphy2024}.   
High-precision, almost continuous, photometry with 2-min cadence was obtained by the TESS space mission in Sector 11 (S11; 2019 April\,26 -- May 20; 24\,d time span) and S38 (2021 April\,29 -- May 26; 26\,d time span). 
\cite{Pigulski2022} analyzed 
those data of the HD\,133729 BLAP by using the traditional O$-$C method, where only the times of the maximum light of the pulsation were analyzed. 

In this paper, we apply the FM (frequency modulation) method (\citealt[][hereafter, FM1 and FM2, respectively)]{FM1, FM2}. 
The FM method is advantageous over traditional methods as, through the Fourier transform, it uses all the photometric data to more stringently derive the orbital parameters and a mass constraint,  
while a traditional O$-$C method uses only the times of the pulsation amplitude maximum.
In the present case, the number of the data points in the O$-$C method is $\sim$$10^3$. On the other hand, the FM  method uses all the photometry taken at 2 minute cadence over S11 and S38, so the number of the data points is a factor $\sim$$15$ larger than the O$-$C method. Hence the FM method leads to more accurate results.

\section{Fourier analysis of \\the TESS photometry data}
As demonstrated in FM2, Fourier analysis of the luminosity variation of a pulsating star in a binary system leads to multiplets in the amplitude spectrum with frequency separation between adjacent peaks equal to the orbital frequency. 
The luminosity variation of HD\,133729 has a sawtooth shape, giving rise to a harmonic series in the amplitude spectrum, with each harmonic showing a multiplet structure.
Using S11 and S38 we performed a Fourier analysis of the main peak of the harmonic series by using {\sc period04} \citep{Period04} and the Discrete Fourier transform of \citet{1985MNRAS.213..773K}. 
We extracted the central quintuplet of frequencies and ignored the lower amplitude, higher order harmonics.
We fitted those five frequencies by linear least-squares to the data, choosing  $t_0$ to make the phases of the first sidelobes equal. 
The results are shown in Table\,\ref{tab:01}. The frequency of the central peak, $\nu_0$, is the frequency of the pulsation mode, which is likely to be the radial fundamental mode. As seen in Table\,\ref{tab:01}, the first and the second sidelobes are separated from the central peak by exactly $\nu_{\rm orb}$ and $2\nu_{\rm orb}$, respectively.  
Hereafter, the amplitudes and the phases of sidelobes at $\nu_0 + m\nu_{\rm orb}$ ($m=0, \pm 1, \pm 2$) are denoted by $A_m$ and $\theta_m$, respectively.

\begin{deluxetable*}{lclrlr}
\tablenum{1}
\tablecaption{A least-squares fit of the frequency quintuplet to the  S11 and S38 TESS data\label{tab:01}; $t_0 = {\rm BJD}\,2458978.02692$. Note that the phase difference between that of the central frequency and the first sidelobes is equal to $\uppi/2$\,rad, the signature of pure frequency modulation, as expected for FM. }
\tablewidth{0pt}
\tablehead{
 \multicolumn{2}{c}{frequency (d$^{-1}$)} & 
 \multicolumn{2}{c}{amplitude (mmag)} & 
 \multicolumn{2}{c}{phase (rad)}
 }
\startdata
$\nu_0-2\nu_{\rm orb}$ & $44.400101$ & $A_{-2}$ & $0.204 \pm 0.007$ & $\theta_{-2}$ & $-0.0474 \pm 0.0069$ \\
$\nu_0-\nu_{\rm orb}$ & $44.443420$ & $A_{-1}$ & $1.841 \pm 0.007$ & $\theta_{-1}$ & $1.5442 \pm 0.0008$ \\
$\nu_0$ & $44.486740$ & $A_{0}$ & $12.857 \pm 0.007$ & $\theta_{0}$ & $3.1294 \pm 0.0001$ \\
$\nu_0+\nu_{\rm orb}$ & $44.530059$ & $A_{+1}$ & $1.846 \pm 0.007$ & $\theta_{+1}$ & $1.5442 \pm 0.0008$ \\
$\nu_0+2\nu_{\rm orb}$ & $44.573379$ & $A_{+2}$ & $0.089 \pm 0.007$ & $\theta_{+2}$ & $-0.0887 \pm 0.0158$ 
\enddata
\end{deluxetable*}

\section{Road map to the FM method {and Results}}
The difference in the light arrival time compared to the the case of a signal arriving from the barycenter of the binary system is expressed in terms of a series expansion of trigonometric functions of the orbital frequency $\nu_{\rm orb} t$, where $t$ is the time after the star passed the periapsis. The coefficient amplitude normalized by the light travel time across the projected semi-major axis $\xi_n$ and the phase $\vartheta_n$ of the $n$-th term ($n=1, 2, \cdots$) of the series expansion are dependent on the eccentricity $e$ and the angle between the periapsis and the nodal point $\varpi$, and their functional forms are explicitly given by equations (14) and (15) of FM2.
The ratio between the light travel time across the projected semi-major axis ($a_{\rm BLAP}\sin i$) 
and the pulsation period of the mode in consideration measures the amplitude of the phase modulation, and it is denoted $\alpha$ (equation (20) of FM2).
 
The FM method utilizes the analytical relations, derived in FM1 and FM2, between the Fourier amplitudes $A_m$ of the photometry data and the amplitudes $\xi_n$ and phases $\vartheta_n$, to derive the orbital parameters.
In short,
\begin{itemize}
\item 
the ratio between the sum of the amplitudes of the first sidelobes to the central peak amplitude, $\left(A_{+1}+A_{-1}\right)/A_0$, is regarded as a function of $\alpha\xi_1$ (equation (30) of FM2),
\item
the ratio between the sum of the amplitudes of the second sidelobes to the central peak amplitude, $\left(A_{+2}+A_{-2}\right)/A_0$, is regarded as a function of $\alpha\xi_2$ (equation (36) of FM2),
\item
the relative difference between the pair of first sidelobes, $\left(A_{+1}-A_{-1}\right)/\left(A_{+1}+A_{-1}\right)$ is regarded as a function of $2\vartheta_1 - \vartheta_2$ and $\alpha\xi_2$ (equation of (31) of FM2).
\end{itemize}

It is the FM technique that makes the best use of these relations.
The road map is divided into several steps:
\begin{enumerate}
\item evaluate  $\left(A_{+1}+A_{-1}\right)/A_0$, $\left(A_{+2}+A_{-2}\right)/A_0$ and $\left(A_{+1}-A_{-1}\right)/\left(A_{+1}+A_{-1}\right)$,
\item determine $\alpha\xi_1$, $\alpha\xi_2$ and $\left(2\vartheta_1-\vartheta_2\right)$, and then $\xi_2/\xi_1$,
\item determine $(e, \varpi)$ from $\left(\xi_2/\xi_1\right)$ and $\left(2\vartheta_1-\vartheta_2\right)$,
\item evaluate $\xi_1$ by substituting $e$ and $\varpi$ thus determined,
\item determine $\alpha$ by using $\alpha\xi_1$ and $\xi_1$,
\item determine $a_{\rm BLAP}\sin i/c$ from $\alpha$,
\item evaluate the mass function $f$ by using $\alpha, \nu_{\rm orb}$ and $\nu_{\rm osc}$ (equation (55) of FM2).
\end{enumerate}

The results derived in this way are as follows: 
\[
\left\{
\begin{array}{lcl}
a_{\rm BLAP}\sin i/c&=&88.7\pm 0.1\,{\rm s}\\
e&=&0.163\pm 0.007\\
\varpi&=&1.70\pm 0.18\,{\rm rad}\\
\nu_{\rm orb}&=&0.0433194\pm 0.0000004\,{\rm d}^{-1}\\
f&=&1.400\pm 0.004\,{\rm M}_\odot\\
\end{array}
\right.
\]

Compared with the O$-$C analysis of \cite{Pigulski2022}, the eccentricity is significantly different. 
We suggest that the main cause of this is the FM method uses the entire photometric data set as opposed to the the O$-$C method which uses only the times of the pulsation amplitude maxima. Also, fitting other long-term trends to the data carried out in the O$-$C process could be the reason, while the FM method does not perform such a fitting procedure.

\section{Discussion}
\cite{Pigulski2022} estimated, from spectral energy distribution analysis, the effective temperature and the surface gravity of the companion of the BLAP of HD\,133729, a B main sequence star, to be $T_{{\rm eff}, {\rm MS}} = 11\,500 \pm 1\,000\,{\rm K}$ and $\log (g_{\rm MS}/ 1\,{\rm cm}\,{\rm s}^{-2})=4.0$, respectively, and 
those of the BLAP itself to be $T_{{\rm eff}, {\rm BLAP}} = 29\,000\pm 1\,000\,{\rm K}$ and $\log (g_{\rm BLAP}/ 1\,{\rm cm}\,{\rm s}^{-2}) = 4.5$, respectively. Along with their inference of the luminosity of the BLAP $\log (L_{\rm BLAP}/{\rm L}_\odot) = 2.15\pm 0.15$, these estimates lead to 
$M_{\rm BLAP}/{\rm M}_\odot = 0.260^{+0.118}_{-0.080}$ and  
$\log (\overline{\rho}_{\rm BLAP}/\overline{\rho}_\odot) = 0.397 \pm 0.081$. Combining these with the pulsation frequency 
$\nu_0 = 44.4867\,{\rm d}^{-1}$, we then get
the `pulsation constant' $Q = P{\left(\overline{\rho}/\overline{\rho}_{\odot}\right)^{1/2}} = 0.036 \pm 0.003$,
which seems reasonable for fundamental mode pulsation in a pre-white dwarf star \citep{Cox1980}.
The other way around, with the reasonable $Q$-value for pre-white dwarfs, the inferred mass of $0.26\,{\rm M}_\odot$ is admissible.

The mass of the companion of the BLAP is estimated, again according to \cite{Pigulski2022}, to be $M_{\rm MS} = 2.85\pm 0.25\,{{\rm M}_\odot}$. It is noteworthy that the implied mass ratio, $q = 0.260/2.85 = 0.091$, is discrepant with a recent binary evolution model of the system where $q=0.30$ was preferred \citep{zhangetal2025}.
From the mass function  $f = 1.400\pm 0.004\,{\rm M}_\odot$ we derived above, which is in reasonable agreement with \cite{Pigulski2022}, we get the most likely value of $\sin i = 0.836$, hence $i=57^\circ$.

By using the orbital parameters, we may deduce the radial velocity curves of both the BLAP and the B main sequence star (equation (13) of FM2). With spectroscopic measurement of radial velocities, the present FM results could be independently verified. The radial velocity amplitude of the B main sequence component is of the order of $\sim 8\,{\rm km}\,{\rm s}^{-1}$, which is measurable.

\bibliographystyle{aasjournal}
\bibliography{references_BLAP}  

\end{document}